\documentclass[a4paper]{article}
 \usepackage{graphicx}
 \usepackage{amsmath}
 \usepackage{hyperref}
 \usepackage{color}

\begin{document}
 
 \begin{center}
 
 {\bf \Large Deterministic ants in labirynth - information gained by map sharing }\\[5mm]

 {\large Janusz Malinowski and Krzysztof Ku{\l}akowski}\\[3mm]

 {\em
 
 Faculty of Physics and Applied Computer Science, AGH University of
 Science and Technology, al. Mickiewicza 30, PL-30059 Krak\'ow,
 Poland\\

 }

 
 {\tt kulakowski@fis.agh.edu.pl}

 \bigskip
 
 \today
 
 \end{center}
 
 \begin{abstract}

A few of ant robots are dropped to a labirynth, formed by a square lattice with a small number of nodes removed. Ants move according to a deterministic algorithm designed to explore all corridors. Each ant remembers the shape of corridors which she has visited. Once two ants met, they share the information acquired. We evaluate how the time of getting a complete information by an ant depends on the number of ants, and how the length known by an ant depends on time. Numerical results are presented in the form of scaling relations.
\end{abstract}
 
 \noindent
 
 
 \noindent
 

 \bigskip
 \section{Introduction}

In 2003, Ong and coworkers presented a vision of history of man-machine interaction \cite{ong03}. According to this vision, the 
interaction evolves from 'manual operation' through 'machine assisted operation' and 'remote operations in- and out-of-sight' to 
'multiple robot management' and 'unsupervised autonomous robot management'. At the last stage of this story, 'multiple unsupervised 
autonomous robots management' is expected. As it was stated two years ago by the same authors, we are at the fourth level: 
'remote operation within area of sight' \cite{wong11}. Then it would be timely to think about 'multiple robot management'. 
On the other hand, cooperation between robots is of interest for tens of years, with the ant colony optimization algorithm 
as a milestone \cite{dori}. More recent works are at the level of applications (\cite{dada,suwon,soccer} to call only a few). 
In particular, cooperation between Search-and-Rescue (SAR) robots is of current interest \cite{wang08,vis08}. Urban SAR robots 
(\cite{bala06,vela08,wang08,vela08a,sun08,pfing08,vis08,vis08a} and references therein) can be useful in emergency situations, like 
earthquakes, urban fires or battlefields, for victim localization. In \cite{vis08}, efficiency of the penetration of a disaster area
was evaluated via entropy minimization, performed by two robots; the entropy was due to the unknown spatial distribution 
of obstacles. Knowledge on this distribution was equivalent to a map.\\ 

Our formulation of the research question is related to a map sharing. We assume that a robot ant can efficiently collect 
an information on the corridors she visited. Further, once two ants met, they can exchange the gathered information. If there is only 
one ant and she moves in a random way without differentiating between known and unknown corridors, the radius $R$ of the known area 
is expected to vary as square root of time. Using the parameterization 
$R\propto t^{\alpha}$, we get the exponent $\alpha = 1/2$; this is the case of normal diffusion. Surprisingly enough, the same value of $\alpha$ 
can be produced by purely deterministic rules of motion; this is the so-called deterministic diffusion \cite{szuster}. On the contrary,
the condition of self-avoiding produces an increase of the exponent in less-than-four-dimensional systems; the Flory formula 
$\alpha=3/(d+2)$, where $d$ is the dimensionality, gives results reasonably close to numerical evaluations \cite{note}. 
This is the case of hyperdiffusion \cite{note,buszo}.  (The values of $\alpha$ smaller than 1/2 are classified as subdiffusion; both hyper- and sub-diffusion are termed 'anomalous diffusion'.) For $d \ge 4$, the exponent $\alpha$ is 1/2 again; the condition of self-avoiding is not demanding there.\\

The value of the exponent $\alpha$ is relevant for an evaluation of the time $T$, when most ants know the whole labirynth. Suppose for a moment, that
the area penetrated by one ant till time $t$ is $t^\alpha$. Naively, ants met and exchange information at time $T$ when $NT^\alpha$ covers the whole area; here $N$ is the number of ants. This evaluation suggests that $T\propto N^{-1/\alpha}$. Here we assume that ants move according to a deterministic algorithm designed as to penetrate the whole labirynth. Then, an ant tries to omit the places she already knows from her previous path or from other ants. We ask, what is the relation between $T$ and $N$? Also, we are interested also in the time dependence of the number $n(t)$ of known cells.\\

Below we distinguish between the theoretical exponent $\alpha$ related, as above, with the diffusion theory, the exponent $\beta$ related to the time $T$ of penetration of the whole labirynth as dependent on the number of ants $N$, and the exponent $\gamma$ related to the time dependence of the number of visited cells. Both $\beta$ and $\gamma$ are obtained numerically. Additionally, the latter exponent is calculated for the case of random walk; this will be termed $\gamma '$.\\

In our modeling, the environment is represented by a square lattice with some small amount ($p$ percent) of nodes removed; hence labirynth 
in the title. The rules of ant motion are designed as to assure that the whole environment will eventually be penetrated. These rules are
explained in detail in Section 2.   The numerical results are given in Section 3 in the form of scaling relation, as the one between $n$ and $t$.
Scaling relations are useful if there is no characteristic scale within the investigated range of parameters. In our case, we can expect 
different behaviour before and after ants meet; therefore the scaling exponents can be different for short and long time. Last section 
is devoted to conclusions.

\section{Algorithm and calculations}

The labirynth is created as a square lattice 0f $97\times 57 = 4185$ nodes, arranged as in Fig. 1. Next, a small amount of randomly selected nodes ($p$ percent) are removed from the lattice. As $p$ is not greater than ten percent, what is less than the percolation threshold, the lattice is expected to be connected \cite{sta,oth}. The accessibility of nodes is verified by the standard flood fill algorithm \cite{flofi}; if after the removal of a node some other area cannot be accessed, the node cannot be removed. We have checked that if an ant moves randomly in an undamaged lattice, the mean squared distance from the beginning of the trajectory increases at early times as $t^{0.989}$, in accordance with standard theory of random walk. \\

The algorithm to penetrate the labirynth is constructed as follows. An ant leaves seeds at visited nodes. If she is at a dead end (degree one), she leaves two seeds there. If she is at a corridor (degree two), she leaves one seed. If she has to go back in this corridor, she leaves two seeds there.
If she is at a node of degree 3 or 4, she checks the number $S$ of seeds present at neighboring nodes, finds a node with minimal $S$, goes there and leaves one seed there. If two neighboring nodes are of equal $S$, a half of ants selects a first node from the left; another half - from the right.
Then, there are two kinds of ants in the labirynth. In all investigated cases, this algorithm allows to penetrate the whole labirynth.\\

Each point for the curves $T(N)$ is obtained for 500 different runs. Each run is a new labirynth (except $p=0$, where there is no removed nodes) and new initial positions of ants. The curves $n(t)$ are averaged over 30 ants.

\section{Results}

 \begin{figure}[ht]
 \centering {\centering \resizebox*{13cm}{10cm}{\rotatebox{-00}{\includegraphics{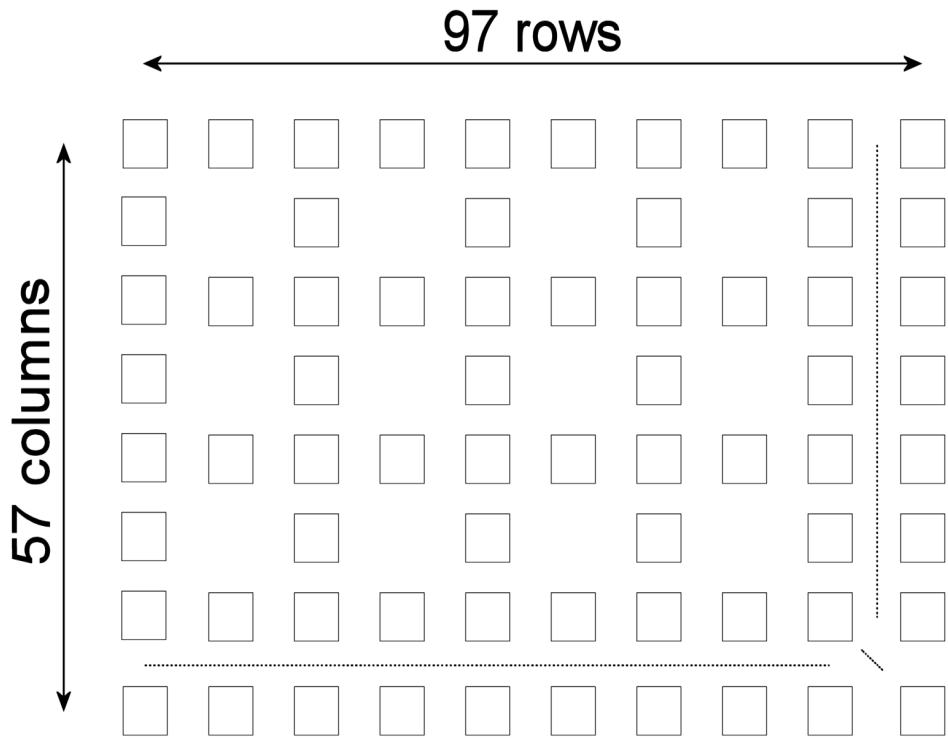}}}}
\caption{A part of the labirynth.}
\label{fig-1}
\end{figure}

 \begin{figure}[ht]
 \centering {\centering \resizebox*{13cm}{10cm}{\rotatebox{-00}{\includegraphics{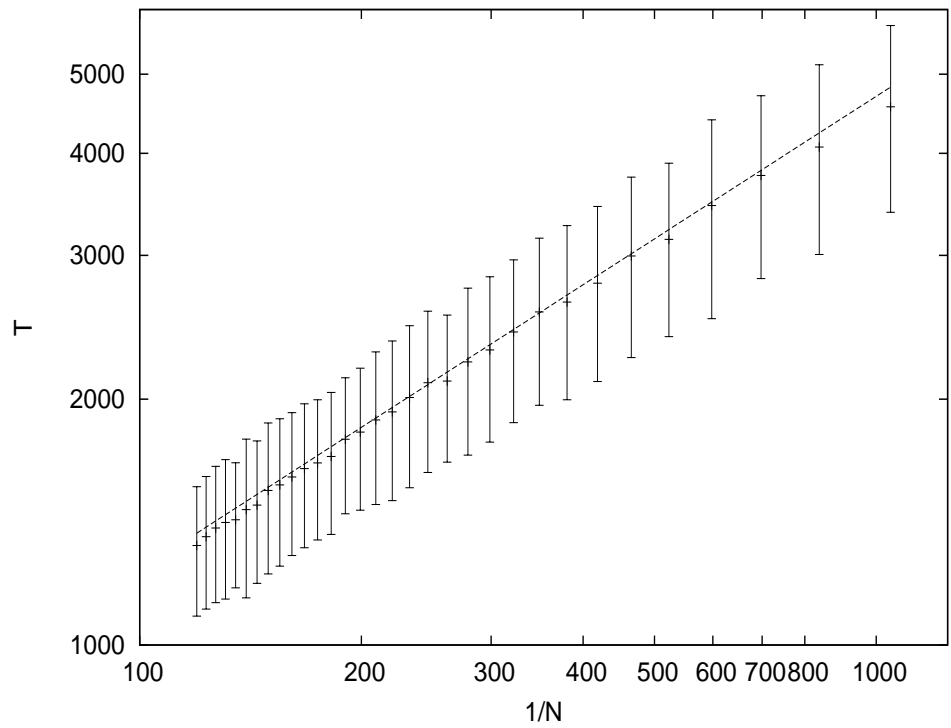}}}}
 \centering {\centering \resizebox*{13cm}{10cm}{\rotatebox{-00}{\includegraphics{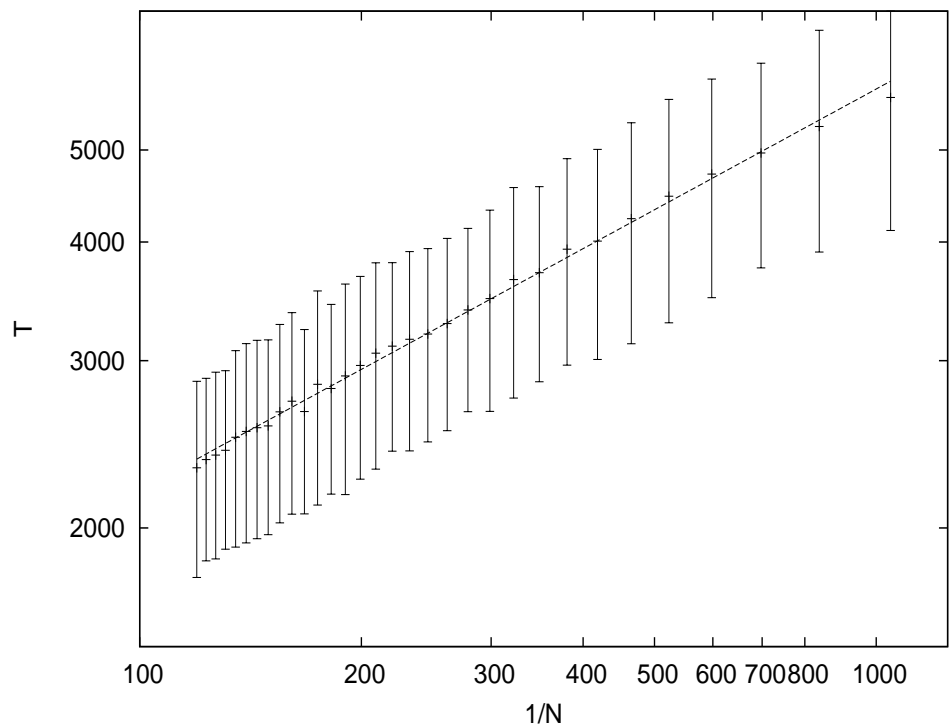}}}}
\caption{The time $T$ needed to penetrate the whole labirynth by (a) first ant, (b) last ant, as dependent on the number $N$ of ants. Each point is an average over 500 different runs. Each run is a new labirynth and initial positions of ants. }
\label{fig-2}
\end{figure}

 \begin{figure}[ht]
 \centering {\centering \resizebox*{13cm}{10cm}{\rotatebox{-00}{\includegraphics{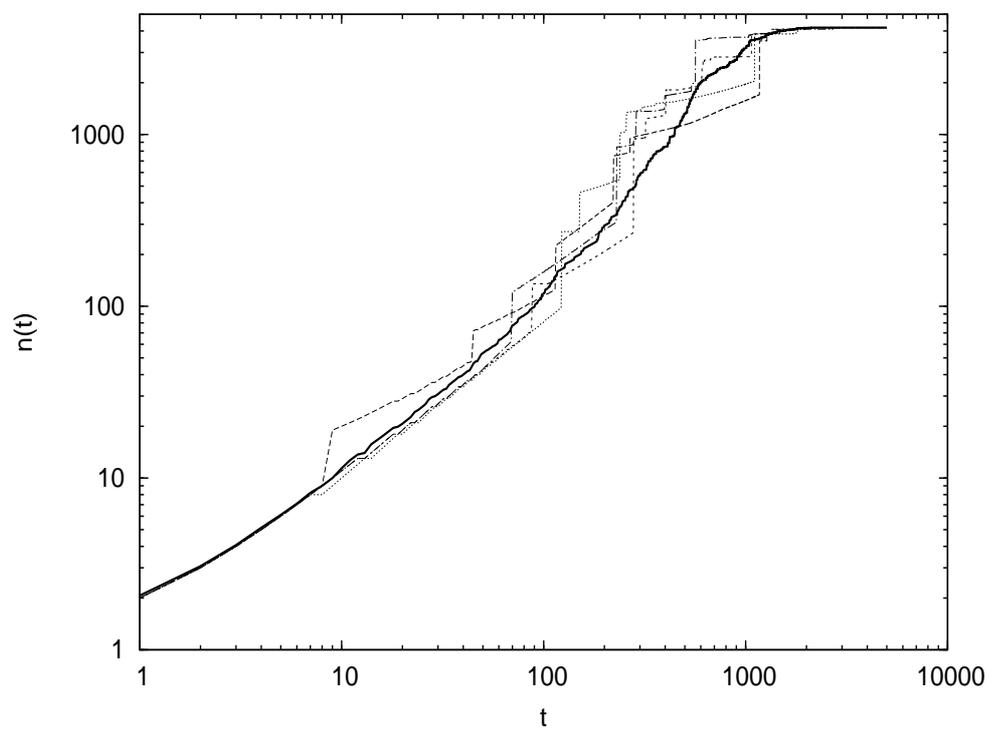}}}}
\caption{Time dependence of the number of cells, known by a deterministic ant. Three dotted curves are related to three different ants, and the continuous line - to the average over 30 ants.}
\label{fig-3}
\end{figure}

 \begin{figure}[ht]
 \centering {\centering \resizebox*{13cm}{10cm}{\rotatebox{-00}{\includegraphics{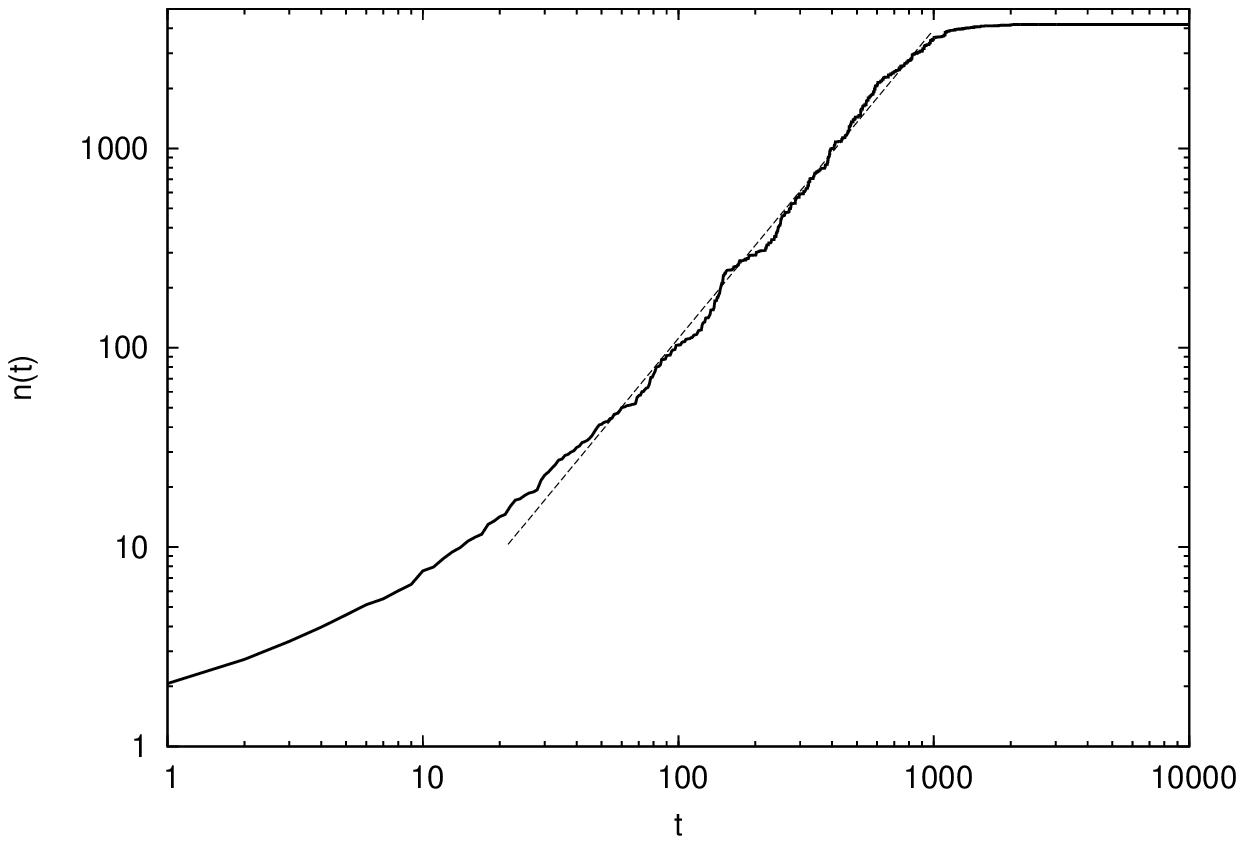}}}}
\caption{Time dependence of the number of cells, known by a random walker. The average over 30 ants is shown with a fit within the limited time range.}
\label{fig-4}
\end{figure}

The relations between the time of penetration of the whole labirynth and the number of ants are obtained numerically for two cases: \\
a) first ant knows the whole labirynth,\\
b) last ant (what is equivalent to all ants) knows the whole labirynth.\\

An exemplary dependence of $T(N)$ is shown in Fig. 2. The obtained exponents $\beta$ in the relation $T\propto N^{-\beta}$ are different in the case (a) and (b). They are also different
for $p=0$ and $p>0$. For a complete lattice ($p=0$), we get $\beta$ equal to (a) $0.58\pm 0.004$ , and (b) $0.42\pm0.005$. For an incomplete lattice, the results are obtained for (a) $p =$ 0.005, 0.01, 0.015, 0.02, 0.025, 0.03, 0.035, 0.04, 0.045, 0.05, 0.075, 0.01, 0.125 and 0.15, and (b) 0.025, 
0.05, 0.075, 0.1, 0.125 and 0.15. The results on $\beta$ fluctuate between (a) 0.67 and 0.69 ($\pm 0.01$), and (b) 0.478 and 0.481 ($\pm 0.009$ or less).
We have no numerical indications that there is any dependence of $\beta$ on $p$, once $p>0$.\\

These results do not distinguish between different stages of the simulation. More insight can be obtained from the time dependence of the 
number of visited cells, $n(t)$, which we intend to fit by the scaling relation $t^\gamma$. However, numerical results indicate that this fit works only in some range of time. The limitation of this range from above is an obvious consequence of the finiteness of the labirynth. The limitation from below can be due to the characteristic time when ants meet. \\

Our numerical results indicate that the exponent $\gamma$ clearly increases with the number of ants. The dependence of $\gamma$ on 
the fraction $p$ of removed nodes is rather weak. The data are presented in Table 1. There, the number of digits shows the numerical accuracy, taken as the mean square root between the simulated curves. In parenthesis, the range of time is added where the obtained function was fitted. Exemplary curves are shown in Fig. 3. \\

\begin{table}
\caption{The exponent Gamma}
\centering
\begin{tabular}{c c c c}
\hline
N    & p=0    & p=0.05  & p=0.1\\
1    & 0.97 (100-5000) & 0.95 (100-5000)   & 0.94 (100-5000)\\
10   & 1.33 (100-1000)  & 1.44 (100-800)   & 1.35 (200-1000)\\
20   & 1.59 (100-1000)  & 1.54 (200-700)    & 1.48 (200-1000)\\
50   & 1.75 (100-500)    & 1.61 (100-500)    & 1.52 (100-600) \\
100  & 1.82 (30-300)  & 1.74 (100-400)    & 1.60 (100-400)\\
\hline
\end{tabular}
\label{gamma}
\end{table}

For the sake of comparison, we calculated also the exponent $\gamma '$ for the case when ants perform random walk and communicate. These results are gathered in Table 2. In parenthesis, the range of time is added where the obtained function was fitted. An example of fit is shown in Fig. 4. \\
 
\begin{table}
\caption{The exponent Gamma'}
\centering
\begin{tabular}{c c c c}
\hline
N    & p=0    & p=0.05  & p=0.1\\
1    & 0.8111 (10-11000) & 0.772 (20-20000)   & 0.745 (20-20000)\\
10   & 1.051 (10-8000) & 1.039 (10-8000) & 1.090 (10-8000)\\
20   & 1.219 (10-4000)   & 1.205 (10-4000)  & 1.186 (10-4000)\\
50   & 1.376 (10 -1000)   & 1.323 (10-2000)    & 1.325 (10-2000)\\
100  & 1.549 (20-1000) & 1.404 (10-1000)    & 1.578 (100-1000)\\
\hline
\end{tabular}
\label{gammaprim}
\end{table}

As seen in the curves shown in Figs. 3 and 4, at early times $n(t)$ depends on time more mildly. The observed crossover between early and late times is due to the difference between the early stage, where ants do not met yet, and the later stage where ants do exchange information. We also underline that the exponents $\gamma, \gamma '$ are obtained from the last parts of the plots before the plateau. On the other hand, at these parts the curves $n(t)$ seem to be more sharp for longer times.  This means, that if the labirynth size is larger, the obtained values of the exponents can be larger; therefore the obtained values can be treated as lower bounds of the exponents. 

\section{Discussion}

The differences between the obtained values of the exponent $\beta$ for $p=0$ and $p>0$ can be understood in terms of symmetry of the lattice, which is higher when there is no removed nodes. Similar effect is known to influence scaling relations in fractals (\cite{note}, p. 60). For practical applications, the results for $p>0$ are certainly more appropriate, because buildings of high symmetry are rather rare. The obtained values of the exponent $\beta$ are the main results of this work. They can be helpful when evaluating how the search time depends on the number of searchers. We note that the exchange of information is practically important also because it assures that ants do not penetrate the area known already by another ant. \\

The comparison of the results on $n(t)$ allows to state that as a rule, $\gamma$ is slightly larger than $\gamma'$. This result indicates that the 
deterministic walk is more efficient that the random walk; this is not a surprise. Also, the exponent $\gamma$ increases with the number $N$ of ants.
This variation reflects the efficiency of communication between ants. The same variation is observed also for the case of random walk. We note however,
that the fitting of the results on $n(t)$ with the scaling relation is not successful. The problem of how the number of visited sites depends on time in random walk on a two-dimensional lattice was discussed in \cite{husz}. The theoretical formula obtained there is not a scaling relation, and therefore the exponents $\gamma, \gamma '$ have no theoretical importance. For labirynths of larger size, the values of these exponents could be even larger, hence they serve here as lower bounds. The obtained dependence of $\gamma(N), \gamma '(N)$ indicate that these exponents are not universal.\\

To conclude, we investigated the problem how exchange of information on the geometry of environment between robots can accelerate the time $T$ necessary 
to penetrate a labirynth, i.e. a connected system of corridors. The obtained $T$ dependence on the number of robots are: $T\propto N^{-0.68}$ for the first robot, and $T\propto N^{-0.48}$ for the last one. These rules can be useful in practical applications, as localization of victims in complex environment, perhaps after some disasters. In less developed technological applications, the ants can represent personal devices, which are capable to register the map of a local environment along the owner's trajectory and transmit it to another device.

 \section*{Acknowledgements} The research is partially supported within the FP7 project SOCIONICAL, No. 231288.


\begin{thebibliography}{99}

\bibitem{ong03} K. W. Ong, G. Seet and S. K. Sim, {\it A Hierarchical Mixed Multi-Agent System}, 
The 4th IEEE International Conference on Control and Automation, 2003, pp. 644 - 648, Montreal, Canada.

\bibitem{wong11} C. Y. Wong, G. Seet and s. K. Sim, {\it Multiple-Robot Systems for USAR: Key
Design Attributes and Deployment Issues}, International Journal of Advanced Robotic Systems, 
8 (2011) 85-101.

\bibitem{dori} M. Dorigo, {\it Optimization, Learning and Natural Algorithms}, PhD thesis, Politecnico di Milano, 1992.

\bibitem{dada} Dandan Zhang, Junzhi Yu and Long Wang, {\it Adaptive task assignment for multiple mobile robots via swarm 
intelligence approach}, Robotics and Autonomous Systems, 55 (2007) 572-588.

\bibitem{suwon} Jin Hong Jung, Sujin Park and Seong-Lyun Kim, {\it Multi-robot path finding with wireless multihop 
communications} Communications Magazine, IEEE, 48 (2010) 126-132.

\bibitem{soccer} F. Santos, L. Almeida, L. S. Lopes, J. L. Azevedo and M. B. Cunha, {\it Communicating among robots in the 
RoboCup Middle-Size League}, LNCS 5949 (2010) 320-331.

\bibitem{bala06} S. Balakirsky, C. Scrapper, S. Carpin, M. Lewis (2006), {\it USARSim: providing a framework for multi-robot performance evaluation}, Proceedings of PerMIS 2006 

\bibitem{vela08} P. Velagapudi, P. Scerri, K. Sycara, H. Wang, M. Lewis, and J. Wang (2008), {\it Scaling Effects in Multi-robot Control}, IEEE/RSJ International Conference on Intelligent Robots and Systems, 22-26 Sept. 2008 Page(s): 2121 - 2126 

\bibitem{wang08} J. Wang, H. Wang, M. Lewis, P. Scerri, P. Velagapudi, and K. Sycara (2008) {\it Experiments in Coordination Demand for MultiRobot Systems}, Proceedings of IEEE SMC International Conference on Distributed Human-Machine Systems (DHMS'08), IEEE 

\bibitem{vela08a} P. Velagapudi, J. Wang, H. Wang, P. Scerri, M. Lewis and K. Sycara (2008), {\it Synchronous vs. Asynchronous Video in Multi-Robot Search}, Proceedings of first International Conference on Advances in Computer-Human Interaction (ACHI'08), IEEE 

\bibitem{sun08} D. Sun, A. Kleiner, and T. M. Wendt (2008), {\it Multi-Robot Range-Only SLAM by Active Sensor Nodes for Urban Search and Rescue}, in  Robocup 2008: Robot Soccer World Cup XII, 2008 

\bibitem{pfing08} M. Pfingsthorn, B. Slamet and A. Visser (2008), {\it A Scalable Hybrid Multi-Robot SLAM Method for Highly Detailed Maps}, Lecture Notes on Artificial Intelligence series, volume 5001, p. 457-464, Springer, Berlin Heidelberg New York, July 2008

\bibitem{vis08} A. Visser and B.A. Slamet (2008), {\it Including communication success in the estimation of information gain for multi-robot exploration}, in "International Workshop on Wireless Multihop Communications in Networked Robotics", Berlin, April 4, 2008.

\bibitem{vis08a} A. Visser and B.A. Slamet (2008) {\it Balancing the information gain against the movement cost for multi-robot frontier exploration}, Springer Tracts in Advanced Robotics, 44 (2008) 43-52. 

\bibitem{szuster} H. G. Schuster and W. Just, {\it Deterministic Chaos. An Introduction}, Wiley-VCH, Weinheim 2005

\bibitem{note} D. Ben-Avraham and S. Havlin, {\it Diffusion and Reactions in Fractals and Disordered Systems} Cambridge UP, Cambridge 2000.

\bibitem{buszo}  J.-P. Bouchaud, {\it Anomalous diffusion in disordered media}, Physics Reports 195 (1990) 127-293.

\bibitem{sta} D. Stauffer and A. Aharony, {\it Introduction to Percolation Theory}, (2nd Ed.), Taylor and Francis, Philadelphia 1991.

\bibitem{oth} V.A. Cherkasova, Yu.Yu. Tarasevich, N.I. Lebovka and N.V. Vygornitskii, {\it Percolation of the aligned dimers on a square lattice}, 
Eur. Phys. J. B 74 (2010) 205-209.

\bibitem{flofi} J. D. Foley and A. Van Dam, {\it Fundamentals of interactive computer graphics}, Reading, Mass., Addison-Wesley 1982.

\bibitem{husz} F. S. Henyey and V. Seshadri, {\it On the number of distinct sites visited in 2D lattices}, J. Chem. Phys. 76 (1982) 5530-5534.

\end{thebibliography}
 \end{document}